\documentclass[doublecol]{epl2}
\usepackage{amsmath,amssymb,graphicx}

\title{Fluctuations of company yearly profits versus scaled revenue:\\
           Fat tail distribution of L\'evy type}

\shorttitle{Fluctuations of company yearly profits vs scaled revenue}

\author{H.~Eduardo~Roman\inst{1}, Riccardo~A.~Siliprandi\inst{1},
                Christian~Dose\inst{2}, Claudia~Riccardi\inst{1} and Markus~Porto\inst{3}}
		
\shortauthor{H.E.~Roman, R.A.~Siliprandi, C.~Dose, C.~Riccardi and  M.~Porto}

\institute{\inst{1} Dipartimento~di~Fisica, Universit\`a~di~Milano-Bicocca,
                                  Piazza~della~Scienza~3, 20126~Milano, Italy\\
                   \inst{2} Hewlett-Packard, Via~Giuseppe~Di~Vittorio~9,
                                  20063~Cernusco~sul~Naviglio~(MI), Italy\\
	           \inst{3} Institut~f\"ur~Festk\"orperphysik,
                                  Technische~Universit\"at~Darmstadt,
                                  Hochschulstr.~8, 64289~Darmstadt, Germany}

\abstract{
We analyze annual revenues and earnings data for the 500 largest-revenue U.S. companies 
during the period 1954-2007. We find that mean year profits are proportional to mean
year revenues, exception made for few anomalous years, from which we postulate a linear
relation between company expected mean profit and revenue. Mean annual revenues are used 
to scale both company profits and revenues. Annual profit fluctuations are obtained as 
difference between actual annual profit and its expected mean value, scaled by a power 
of the revenue to get a stationary behavior as a function of revenue. We find that
profit fluctuations are broadly distributed having approximate power-law tails with a
L\'evy-type exponent $\alpha \simeq 1.7$, from which we derive the associated break-even 
probability distribution. The predictions are compared with empirical data.}


\begin{document}

\maketitle


Predicting forthcoming year company profit is difficult due to the many unknown
variables determining the actual earnings scenario. This intrinsic uncertainty in
economy's evolution makes earnings forecasts not to be correlated to actual earnings
with the desired accuracy. A consequence is that, often, stocks with highest earnings
forecasts dramatically underperfom those with poor forecasts 
(see e.g. \cite{Hwang/Keil/Smith:2004}). 

Indeed, company earnings may undergo dramatic variations from year-to-year, even over shorter
time scales, leading to huge movements in public company stock (see e.g.\ \cite{Earnings:1986}). 
A less volatile quantity is represented by total company revenue, but also in this case revenue 
variations may yield conspicuous changes in the underlying stock price. Interestingly, the connection
between stock price (i.e.\ market value) and revenue is still surrounded by many open questions 
which are awating for further research (see e.g.\ \cite{VCookJr:2007}). Clearly, the question arises 
of how to estimate in a more realistic fashion profit fluctuations, and therefore attempting to improve 
the accuracy of earnings predictions, the latter being closely related to the issue of 
profitability or break-even point \cite{break-even-analysis}. Several attempts have been made in
order to incorporate a stochastic behavior of profits into the analysis (see e.g.\
\cite{Jaedicke/Robichek:1964,Kim:1973,Adar/Barnea/Lev:1977,Kottas/Lau:1978,YunkerYunker:2003,YunkerSchofield:2005}), 
in which fluctuations are assumed to be normally distributed.

From a fundamental point of view, one may wonder whether the above difficulties can be
mitigated to some extent by modifying the way the problem is approached. In physical
many-body systems for example, a first, realistic solution to a problem can be achieved 
if one resorts to the so-called mean-field approximation, in which a single particle `sees'
an average field due to the remaining particles in the system (see e.g. \cite{Parisi,ChaikinLubensky}). 
Particle-particle correlations and fluctuations of physical quantities can be incorporated into the 
formalism at a later stage once the mean-field solution of the problem is known (see e.g. \cite{Broglia}). 
How can we apply this idea to the study of profit fluctuations of real companies, which can be viewed as 
a many-particle system of interacting economic units? Is it possible to come up with the strong 
fluctuations observed in company profits?
 
Profit fluctuations can be naively evaluated by looking at their relative variations say, from 
year to year. As a matter of fact, however, profit is closely related to revenue and production 
costs (see below) and a different approach based on these interrelations could be explored. 

In this Letter, we suggest that revenue can be taken as the independent, driving variable 
and present a novel analysis of profit fluctuations based on this assumption. We support our 
premises using market data from U.S. companies on an year-to-year basis over a period of 54 years. 
The analysis of the empirical data suggests a form for the expected mean profit, being a function 
of company revenue, with respect to which earnings fluctuations can be determined. The latter turn 
out to be dependent on revenue, suggesting that they are not stationary as a function of revenue. 
Invoking then the condition of stationarity, the fluctuations are scaled by a power of revenue with 
an exponent $\eta$ in the range $1/2<\eta<1$. The probability distribution function of scaled fluctuations 
displays slowly-decaying tails which turn out to be of L\'evy type. A further analysis on the resulting 
break-even point yields a prediction, supported by empirical data, for the probability of profitability, 
enlightening the role that market fluctuations play in the problem.

In the following, we briefly review the standard cost-volume-profit (CVP) analysis, from which we 
derive our main conjectures regarding profit fluctuations. Let us consider a generic (typical) company. 
According to standard CVP analysis \cite{CVPanalysis,CVPanalysisII}, we write the profit $P$ as the 
difference between total revenue $R$ and costs, the latter being the sum of variable costs 
$V_{\rm c}$ and additional (sometimes referred to as fixed) costs $F\ge0$, that is
\begin{equation}
P=R-(V_{\rm c}+F).
\label{eq:Profit}
\end{equation}
Further, we write total revenue as $R=v_{\rm s} n_{\rm s}$, where $v_{\rm s}$ is 
the sale price of sold unit and $n_{\rm s}$ the total number of sold units. Similarly, total
variable costs are written as $V_{\rm c}=v_{\rm c} n_{\rm c}$, where $v_{\rm c}$ is the
cost of produced unit and $n_{\rm c}$ is the total number of produced units.

In what follows, we assume linear relations between sold unit values and produced ones,
according to
\begin{eqnarray}
v_{\rm s}&\simeq& \alpha_{\rm s}~v_{\rm c}, \qquad  1<\alpha_{\rm s}, \nonumber\\
n_{\rm s}&\simeq&  \beta_{\rm s}~n_{\rm c}, \qquad  0\le \beta_{\rm s}\le 1.
\label{eq:linearelations}
\end{eqnarray}
The above assumed linearities do not preclude the coefficients ($\alpha_{\rm s},\beta_{\rm s}$) to 
be time-dependent, similarly to a sort of piecewise linear approximation applied in non-linear CVP 
analysis \cite{CVPnonlinear}. Using these relations in Eq.~(\ref{eq:Profit}), with 
$V_{\rm c}=R/(\alpha_{\rm s}\beta_{\rm s})$, yields
\begin{equation}
P=\gamma_s R-F, \quad {\rm with} \quad 
    \gamma_{\rm s} = {\alpha_{\rm s}\beta_{\rm s}-1\over \alpha_{\rm s}\beta_{\rm s}}.
\label{eq:ProfitRevCost}
\end{equation}

Since we are interested in the typical behavior of companies, we write down the above
$P$-$R$-$F$ relation in terms of its mean values, $P_0=\big<P\big>$, $R_0=\big<R\big>$
and $F_0=\big<F\big>$, representing averages of $P$, $R$ and $F$ over several 
companies for a given time horizon, say a year, i.e.
\begin{equation}
P_0=\big<\gamma_{\rm s}\big> R_0 - F_0.                           
\label{eq:meanProfitRevF}
\end{equation}

\begin{figure}
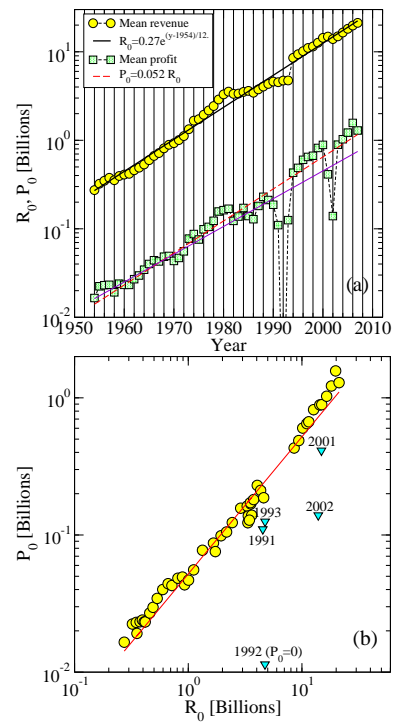

\vspace{0.cm}
\begin{center}
\includegraphics[width=5.cm]{RevProf.eps}
\includegraphics[width=5.cm]{P0R0.eps}
\end{center}
\vspace{-0.5cm}
\caption[]{(color online) \textbf{(a)} Mean yearly revenue, $R_0$ [Billions] (circles) and 
            mean yearly profit, $P_0$ [Billions] (squares) of the 500 largest revenue U.S. 
	    companies \cite{Fortune500} as a function of the year. The thick straight line (top) 
	    is a fit with the form: $R_0=A_0\exp[({\rm year}-1954)/B_0]$ B, with $A_0=0.27$ and 
            $B_0=12$. The dashed line (bottom) is the form: $P_0=A_1\exp[({\rm year}-1954)/B_1]$ B, 
            with $A_1=0.014$ and $B_1=12$. The thin straight line is an exponential regression for
	    $P_0$ over the whole years, yielding: $P_0'=A_2\exp[({\rm year}-1954)/B_2]$ B, 
            with $A_2=0.016$ and $B_2=13.8$, which is not proportional to $R_0$. 
            \textbf{(b)} Mean yearly profit, $P_0$, vs mean yearly revenue, $R_0$ (circles) (from (a)). 
            The straight line is the linear form $P_0=0.052 R_0$. Anomalous-years results are 
            indicated by the down triangles.}
\label{fig:averRev}
\end{figure}

To test this relation, we consider the set of 500 largest revenue companies in the U.S.\ during
the period (1954-2007) \cite{Fortune500}, for which yearly values of $P$ and $R$ are available. 
For each year in the database, we calculate the mean values $P_0$ and $R_0$ in billions (B) of 
U.S.\ dollars. Empirical results for $R_0$ and $P_0$ are plotted in Fig.~\ref{fig:averRev}(a) as 
a function of year. These results suggest that both quantities grow exponentially, and that
$P_0\propto R_0$ if the anomalous years 1991/92/93 and 2001/02 are excluded. The exponential 
dependence of mean profit and revenue also reflects the growth of companies \cite{Stanley:1996}, 
displaying other interesting features described by exponential distribution functions.

Now, to better appreciate the apparent proportionality between mean profits and revenues, we have 
plotted them in Fig.~\ref{fig:averRev}(b), suggesting that indeed
\begin{equation}
P_0\simeq \big<\gamma_{\rm g}\big> R_0, 
\label{eq:meanProfitRev}
\end{equation}
with $\big<\gamma_{\rm g}\big>\cong 0.052$. Slight deviations from linearity can be observed in
Fig.~\ref{fig:averRev}(b) at large revenues, $R_0>10$ B, corresponding to recent last years. This
is an indication that, possibly, non-linear corrections to the result Eq.~(\ref{eq:meanProfitRev})
are playing a role. In keeping with our mean-field strategy, however, we will consider such deviations 
as due to typical market fluctuations. Within this scenario, the model seems to be consistent
with downward profit corrections for 2008, and possibly for the next few years, responding to a sort of 
reverse to the mean linear behavior obtained in Fig.~\ref{fig:averRev}(b). As a matter of fact, there 
is already a widespread consensus that 2008 is manifesting a rather weak economic environment. 
Furthermore, the linear relation between $P_0$ and $R_0$ then suggests that 
\begin{equation}
F_0\simeq \big<\gamma_{\rm c}\big> R_0,
\label{eq:meanCostsRev}
\end{equation}
such that 
\begin{equation}
\big<\gamma_{\rm g}\big>=\big<\gamma_{\rm s}\big>-\big<\gamma_{\rm c}\big>.
\label{eq:meangammag}
\end{equation}
Although we do not know $\big<\gamma_{\rm c}\big>$ explicitely, one can argue that
$0<\big<\gamma_{\rm c}\big><\big<\gamma_{\rm s}\big>$, as one would expect from the definition 
of $F$ (see Eq.~(\ref{eq:Profit})) and the fact that $\big<\gamma_{\rm g}\big>>0$. Now, writing 
$\big<\gamma_{\rm s}\big>=1-\big<1/(\alpha_{\rm s}\beta_{\rm s})\big>$, we can estimate the 
last term by assuming 
$\big<1/(\alpha_{\rm s}\beta_{\rm s})\big>\simeq 1/\big<\alpha_{\rm s}\beta_{\rm s}\big>$.
Accordingly, we find that on average 
\begin{equation}
\big<\alpha_{\rm s}\beta_{\rm s}\big>\simeq\frac{1}{1-(\big<\gamma_{\rm g}\big>+\big<\gamma_{\rm c}\big>)}, 
\end{equation}
hence $\big<\alpha_{\rm s}\beta_{\rm s}\big>>1/(1-\big<\gamma_{\rm g}\big>)\simeq1.062$ and
$\big<\gamma_{\rm s}\big>>0.058$, implying that $\big<\gamma_{\rm c}\big>>0.006$. 

The linear relation between $P_0$ and $R_0$ is obtained when the anomalous years (1991, 1992, 1993, 
2001 and 2002, see Fig.~\ref{fig:averRev}(a)) are excluded from the exponential fit. If these 
anomalous years are included in the fit, an exponential regression for $P_0$ (see the thin straight 
line in Fig.~\ref{fig:averRev}(a)) yields a non-linear relation between $P_0$ and $R_0$.
In the light of these results we may argue that companies show typical profit-revenue scenario
when both variables are linearly related to each other, at least in an average sense. Departures 
from linearity, as the down-triangles shown in Fig.~\ref{fig:averRev}(b), may be referred to as 
extremal, non-typical events. Indeed, the cause for such strong deviations from the linear 
relation between $P_0$ and $R_0$ can be traced back to specific historical facts\footnote{We
refer to the gulf war crisis in 1990-1991 and the resulting economy's recession, and the 
post-internet-bubble- and 9/11-effects during 2001-2002.}.

\begin{figure}
\vspace{0.cm}
\begin{center}
\includegraphics[width=5.cm]{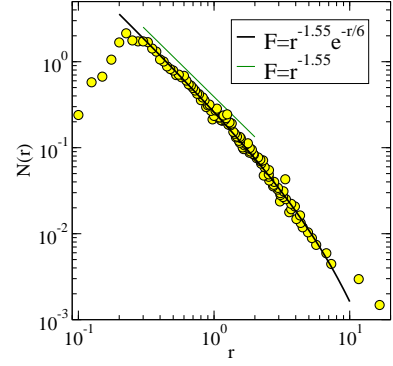}
\end{center}
\vspace{-0.5cm}
\caption[]{(color online) Distribution function of revenues $N(r)$ vs scaled revenue 
            $r=R/R_0$ (circles). The continuous line is a fit with the form: 
	    $n(r)=0.3~r^{-1.55}\exp(-r/6)$ for $r>0.2$. The straight line is the power-law 
	    form: $n(r)\sim r^{-1.55}$ (valid within the interval
	    $0.3\lesssim r \lesssim2$).}
\label{fig:pdfdens}
\end{figure}

The simple exponential fit for $R_0$ shown in Fig.~\ref{fig:averRev}(a) will be used
in the following to scale annual revenues and profits, to take into account the year-to-year 
variations due to the exponential growth in economic activity. One example of such scaled quantities 
is the revenue itself for which we have calculated the probability distribution function (PDF) 
$N(r)$ of scaled revenues $r=R/R_0$. The results are shown in Fig.~\ref{fig:pdfdens}. The scaled 
revenue PDF $N(r)$ displays an intermediate power-law regime with decaying exponent $\simeq -1.55$
followed by an exponential tail for large $r$.  In what follows, scaled profits
will be denoted as $p=P/R_0$.

Next, we study the issue of profit fluctuations by conside\-ring again a generic company having 
scaled profit $p$ and revenue $r$, at any given year within our database. Profit fluctuations
will be considered to be a function of scaled revenue $r$, rather than a function of time. Profit 
fluctuation, denoted as $\Delta p$, is defined as the difference between actual $p$ and its 
expected mean value, here denoted as $\bar{p}$, i.\ e.\
\begin{equation}
\Delta p=p-\bar{p}.
\label{eq:deltaP}
\end{equation}
In order to determine the expected mean profit $\bar{p}$, we have plotted all available values of
$p$ and $r$ in our database (including the anomalous years) and performed a linear regression to the
data which should represent the behavior of $\bar{p}$ vs $r$. The fit, $\bar{p}\simeq a+b r$
(not shown here), yields a rather small intercept value, $a\simeq -0.004$, which can be neglected for our 
present purposes, while $b\simeq 0.056$. Therefore, we postulate that the expected mean profit $\bar{p}$ 
is a function of $r$ and obeying
\begin{equation}
\bar{p}=\big<\gamma_{\rm g}\big> r, \quad {\rm with} \quad \big<\gamma_{\rm g}\big>\cong 0.052.
\label{eq:meanp}
\end{equation}
We will explain below the reason for choosing the above value for $\big<\gamma_{\rm g}\big>$.
Thus, in the present context, company mean profit is a function of solely actual company revenue 
$r$, times a global market parameter $\big<\gamma_{\rm g}\big>$, which is taken the same for all 
companies. Later, we will relax the latter condition and discuss the consequences of taken instead 
a different proportionality factor $\gamma_{\rm g}$ for each individual company. Note that by 
averaging Eq.~(\ref{eq:meanp}) over all companies and years we get 
$\big<\bar{p}\big>\equiv P_0/R_0=\big<\gamma_{\rm g}\big>$, consistent with the empirical result 
Eq.~(\ref{eq:meanProfitRev}). We have also checked that $r\sim 1$ and fluctuations of $r$
are essentially stationary all over the period considered.

The fluctuations Eq.~(\ref{eq:deltaP}) can be scaled by using a characteristic value, such as the 
standard deviation, $\sigma_{\rm p}$, provided that the second moment of the distribution be finite. 
Alternatively, one can use a lower-order moment such as $\big<|\Delta p|\big>$ to characterize the 
amplitude of profit fluctuations.

\begin{figure}
\vspace{0.cm}
\begin{center}
\includegraphics[width=5.cm]{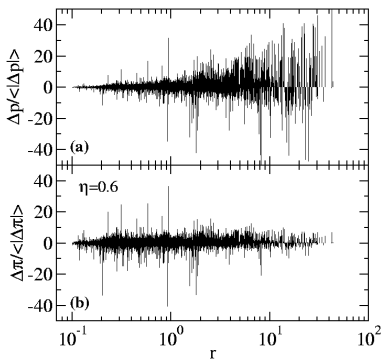}
\end{center}
\vspace{-0.5cm}
\caption[]{(color online) \textbf{(a)} Scaled profit fluctuations $\Delta p/\big<|\Delta p|\big>$ vs 
            scaled revenue $r=R/R_0$, with $\big<|\Delta p|\big>=0.034$.
            \textbf{(b)} Renormalized scaled profit fluctuations $\Delta \pi/\big<|\Delta \pi|\big>$ 
            vs $r$, where $\Delta\pi=\Delta p/r^\eta$, with $\eta=0.6$ and
            $\big<|\Delta \pi|\big>\simeq 0.030$.}
\label{fig:deltaPs}
\end{figure}

Values of $\Delta p$ are plotted in Fig.~\ref{fig:deltaPs}(a) versus scaled revenue $r$.  
As one can see from the plot, profit fluctuations are not `stationary' as a function of $r$,
their amplitudes tend to grow with $r$; the larger the revenue the larger the amplitude of
profit fluctuations. In the following, we wish to quantify the observed rate of growth of amplitude 
fluctuations with revenue and, as a result, being able to find out a source of profit fluctuations 
which is stationary as a function of $r$. To achieve this, we suggest that a suitable variable 
describing fluctuations is given by
\begin{equation}
\Delta\pi = \frac{\Delta p}{r^{\eta}}, 
\label{eq:renorDeltaP}
\end{equation}
with $\eta\ge0$. The question arises of how to determine $\eta$. To do this, we look at the mean-square
fluctuations of the data, $\sigma^2_\eta=\big<(\Delta\pi(\eta))^2\big>-\big<\Delta\pi(\eta)\big>^2$,
for fixed $\eta$, and search for the minimum of $\sigma^2_\eta$ as a function of $\eta$.
We find a minimum value $\sigma_\eta\simeq 0.051$ for $\eta\simeq0.6$. The resulting scaled 
fluctuations $\Delta\pi/\big<|\Delta\pi|\big>$ are reported in Fig.~\ref{fig:deltaPs}(b) versus 
$r$, displaying a satisfactory stationarity. The value of $\eta$ thus obtained does not guarantee 
the vanishing of the first moment $\big<\Delta\pi\big>$. A fine tuning of the value of
$\big<\gamma_{\rm g}\big>$, which enters Eq.~(\ref{eq:deltaP}) and Eq.~(\ref{eq:meanp}), can
be accurately performed in order that $\big<\Delta\pi\big>=0$. This is achieved for
$\big<\gamma_{\rm g}\big>\simeq 0.052$, the value used in the discussions above.

\begin{figure}
\vspace{0.cm}
\begin{center}
\includegraphics[width=5.cm]{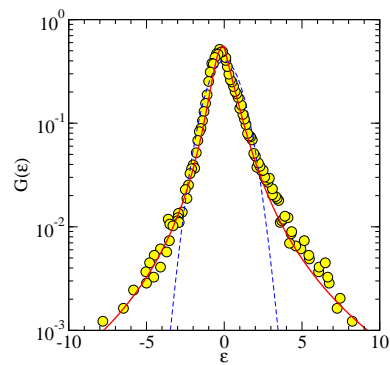}
\end{center}
\vspace{-0.5cm}
\caption[]{(color online) PDF of scaled profit fluctuations $G(\epsilon)$ versus
            $\epsilon=\Delta \pi/\big<|\Delta \pi|\big>$, for  $\eta=0.6$ (circles). The 
	    continuous line is a fit with the form: $F=a/(1+|\epsilon/b|+|\epsilon/c|^{2.7})$,
	    with $a=0.55$, $b=0.6$ and $c=0.9$ for $\epsilon\le0$, and $F=a/(1+|\epsilon/c|^{2.7})$ 
	    with $a=0.55$ and $c=0.75$, for $\epsilon\ge0$, implying power-law tails with a
	    L\'evy-like exponent $\alpha=1.7$. The dashed line is the normal distribution.}
\label{fig:pdfProfit}
\end{figure}

It is remarkable that $\eta\ne 0.5$, the latter would indicate a standard behavior of fluctuations.
The fact that $\eta>1/2$ tells us that fluctuations are stronger than one would expect if $\Delta \pi$
were normally distributed (see e.g.\ \cite{Kim:1973}). To find out the actual shape of the probability 
distribution function, $G(\epsilon)$, for $\epsilon\equiv\Delta \pi/\big<|\Delta \pi|\big>$ (in the case 
$\eta=0.6$), we have plotted it in Fig.~\ref{fig:pdfProfit}. As one can see, the shape of $G(\epsilon)$ 
is consistent with a power-law decay at the tails with a L\'evy-like exponent $\alpha=1.7$. The 
negative tail of $G(\epsilon)$ somehow reflects the fact that companies with poor revenue behavior can be 
taken out of the Fortune 500 set and it may thus become underweighted. Similar arguments are used in discussing 
fund performance (see e.g. \cite{Brown:1992,Elton:1996}) but a thorough understanding of this phenomenon calls
for further studies.
To be noted is that processes resulting from other human-based activity, such as price variations in 
financial markets (see e.g.\ \cite{Stanley:1995}), speed fluctuations of an ensemble of cars in a closed 
circuit traffic \cite{physicaA}, just to name a few examples, also display strongly fluctuating features 
typically characterized by fat-tail distributions. 

We can make contact with the above obtained value of $\eta$ by arguing that indeed $\eta=1/\alpha$.
To see this, let us write the relation $\epsilon\equiv\Delta \pi/\big<|\Delta \pi|\big>$ as follows
\begin{equation}
p=\big<\gamma_{\rm g}\big> r + \big<|\Delta \pi|\big> r^{\eta}\epsilon,
\label{eq:pr_eta_eps}
\end{equation}
suggesting that profit $p$ is driven `deterministically' by the first term proportional to revenue $r$,
plus a second fluctuating term where amplitude fluctuations are also determined by revenue, but to some 
power $\eta$. From a physical point of view, the first term represents a driving or bias field and the 
second one a stochastic part due to an external random force acting on the system (see e.g. \cite{Kampen}). 
Such a model resembles very closely the simple approach due to Bachelier \cite{Bachelier} for describing 
the temporal evolution of stock prices. In our approach, time is replaced by revenue and $\epsilon$ is 
not normally distributed (see also \cite{Hull}).

Now, imagine we can write the factor $r^{\eta}\epsilon$ as a sum over independent, identically distributed
(according to $G(\epsilon)$) L\'evy-like variables $\epsilon_i$, such that
\begin{equation}
r^{\eta}\epsilon = a \sum_{i=1}^{n_r} \epsilon_i,
\label{eq:reta_eps}
\end{equation}
where $a>0$ is a constant and $n_r$ depends on $r$. A similar picture has been used for the description
of seed production of forests \cite{Eisler:2008}. Invoking the stability of L\'evy distributions, 
the above sum is also L\'evy distributed with the same exponent $\alpha$ as the single variable, 
obeying the scaling relation, $\sum_{i=1}^{n_r} \epsilon_i\simeq \epsilon' n_r^{1/\alpha}$ (see e.g.\
\cite{MantegnaStanley}). Even if the random variables $\epsilon$ are not independent, but long-range
autocorrelated, the sum scales as $n_r^{H}$, where the Hurst exponent $H$ is expected to be 
$H\simeq 1/\alpha$ (see e.g.\ \cite{fBm:math}). Details of the corresponding correlations analysis 
will be discussed elsewhere.

\begin{figure}
\vspace{0.cm}
\begin{center}
\includegraphics[width=5.cm]{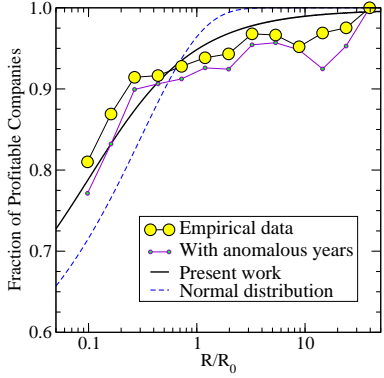}
\end{center}
\vspace{-0.5cm}
\caption[]{(color online) Fraction of profitable (Fortune 500) companies in the U.S. 
            in the period (1954-2007) (circles) versus scaled revenue $r=R/R_0$. The 
            continuous line is the prediction from the present work. The dashed line 
            the one from a normal distribution of profit fluctuations.}
\label{fig:empiricalBE}
\end{figure}

Assuming now that the number $n_r$ is proportional to revenue in the form $n_r=r/r_b$, with $r_b$ 
arbitrarily small, we find according to Eq.~(\ref{eq:reta_eps}), that
\begin{equation}
r^{\eta}\epsilon = a \epsilon' \frac{r^{1/\alpha}}{r_b^{1/\alpha}}.
\label{eq:reta_eps_alpha}
\end{equation}
Identifying the parameter $a$ in the form $a=r_b^{1/\alpha}$ and noting that $\epsilon'\sim\epsilon$, 
we arrive at the relation $\eta=1/\alpha$ claimed above. The values of $\eta$ and $\alpha$ obtained
here are consistent with this prediction.

In what follows, we will elaborate our findings further to consider the issue of profitability,
admittedly important for being able to make predictions about the probability for a generic company
to be profitable. This is related to the concept of break-even or point at which profits vanish. 
The idea here is to estimate the probability of profitability as a function of revenue $r$ by 
appropriately taking into account profit fluctuations due to statistical market variations.

Our derivation starts from Eq.~(\ref{eq:pr_eta_eps}), with $\eta$ substituted by $1/\alpha$, i.e.\
$p=\big<\gamma_{\rm g}\big> r + \big<|\Delta \pi|\big> r^{1/\alpha}\epsilon$, with $\alpha \simeq 1.7$.
At break-even (BE), $p=0$ and the above relation suggests that, 
$\epsilon=-\big<\gamma_{\rm g}\big>r^{1-1/\alpha}/\big<|\Delta \pi|\big>\equiv -\epsilon_{\rm BE}$, where 
\begin{equation}
\epsilon_{\rm BE} = \frac{\big<\gamma_{\rm g}\big>}{\big<|\Delta \pi|\big>} r^{1-1/\alpha},
\label{eq:eps_BE}
\end{equation}
which is positive since here $\big<\gamma_{\rm g}\big> >0$. Now, we define the probability for a generic
company to be profitable, $P_{\rm PF}(r)$, as the fraction of all events for which the fluctuating
variable $\epsilon \ge -\epsilon_{\rm BE}$, since in these cases profit becomes positive, $p>0$.
The probability of profitability can be conveniently expressed as the integral (see also 
\cite{YunkerYunker:2003})
\begin{equation}
P_{\rm PF}(r)= \int_{-\epsilon_{\rm BE}}^{\infty} d\epsilon~G(\epsilon),
\label{eq:P_PF}
\end{equation}
which is a function of $r$ through $\epsilon_{\rm BE}$. One may argue that values $\epsilon\to\infty$ are 
not realistic, requiring the introduction of an effective upper cut-off for $\epsilon$, that we can denote 
as $\epsilon_{\rm cut}$. We find that the results become indistinguishable from those obtained from 
Eq.~(\ref{eq:P_PF}) when $\epsilon_{\rm cut}>8-10$, consistent with the range of variations of $\epsilon$
obtained in Fig.~\ref{fig:pdfProfit}. Thus, for simplicity, we take the upper integration limit in
Eq.~(\ref{eq:P_PF}) as $\epsilon\to\infty$.

\begin{figure}
\vspace{0.cm}
\begin{center}
\includegraphics[width=5.cm]{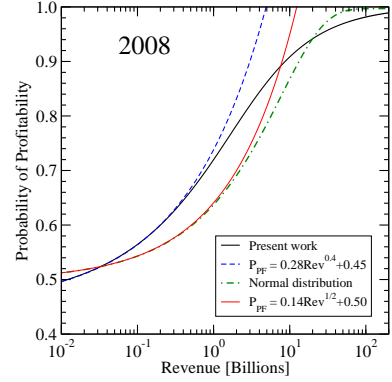}
\end{center}
\vspace{-0.5cm}
\caption[]{(color online) Probability of profitability $P_{\rm PF}$ versus revenue
            [Billions] predicted for 2008 for a typical company. The continuous line is the 
	    theoretical prediction Eq.~(\ref{eq:P_PF}) and the dashed line the linear 
	    approximation valid for small revenues (here $R\lesssim 0.5$ B). 
	    The dot-dashed line is the prediction from the normal distribution and the 
            thin line the corresponding asymptotic linear form.}
\label{fig:breakevenP}
\end{figure}

Numerical results are shown in Fig.~\ref{fig:empiricalBE}, where a comparison is made with the 
total fraction of profitable companies in the period 1954-2007 with respect to the total number
of companies as a function of scaled revenue. The large circles refer to `typical' years, excluding 
the anomalous ones. If the latter are also included, the fraction of profitability decreases a bit. 
Our theoretical prediction works satisfactorily well, justifying a posteriori the few ad-hoc assumptions 
made in this work. The dashed line is the theoretical prediction in the case in which $\eta=1/2$ and 
$G(\epsilon)$ be the normal distribution, clearly yielding a poorer description of the empirical results.

Based on these results, we can make predictions for 2008 using the expected mean revenue
$R_0\simeq 24.3$~B. The results of profitability as a function of absolute revenue $R$ 
(in billions) are displayed in Fig.~\ref{fig:breakevenP}. Linear approximations, valid 
for small revenues, are also reported in the plot to help making simple estimates for 
small revenue companies.

Our results are based on the use of the mean (and positive) growth factor 
$\big<\gamma_{\rm g}\big>$ determining expected profits. This was done as an attempt 
to describe a generic, typical company. Now, actual companies may behave quite differently 
than this typical behavior. This can be taken into account by considering, instead of 
$\big<\gamma_{\rm g}\big>$, the actual company-dependent driving factor 
$\gamma_{\rm g}=\gamma_{\rm s}-\gamma_{\rm c}$, that is the counterpart of Eq.~(\ref{eq:meangammag}),
such that
\begin{equation}
\bar{p} = \gamma_{\rm g} r = (\gamma_{\rm s}-\gamma_{\rm c}) r.
\label{eq:gammacomp}
\end{equation}
Here, the factor $\gamma_{\rm c}$ relates (fixed) costs $F$ to revenue as $F=\gamma_{\rm c}R$, which
in the scaled form becomes $f=F/R_0=\gamma_{\rm c}r$.

The present break-even and profitability results, valid for a typical company, can still be 
applied to a single company with the condition that the break-even value is calculated according 
to the particular value of $\gamma_{\rm g}$. The result is
\begin{equation}
\epsilon_{\rm BE} = \frac{(\gamma_{\rm s}-\gamma_{\rm c})}{\big<|\Delta \pi|\big>} r^{1-1/\alpha},
\label{eq:eps_BE_neg}
\end{equation}
which can become negative if $\gamma_{\rm s}<\gamma_{\rm c}$. In the case $\gamma_{\rm g}>0$, 
also $\epsilon_{\rm BE}>0$, and the previous conclusions for $P_{\rm PF}(r)$ still apply.
When $\gamma_{\rm g}<0$, then $\epsilon_{\rm BE}<0$ and the lower integration limit in 
Eq.~(\ref{eq:P_PF}) becomes positive, yielding lower values of $P_{\rm PF}(r)$ as compared to
$P_{\rm PF}(r)$ with the same revenue $r$ but positive $\gamma_{\rm g}$. Thus, for a single company
the problem reduces to estimate accurately the growth factors $\gamma_{\rm s}$ and $\gamma_{\rm c}$ 
in each particular scenario.

In summary, we have analyzed annual profits and revenues of U.S. companies over a period of 54 years. 
We find a linear relation between annual mean profit ($P_0$) and revenue ($R_0$), which is at the basis 
of the concept of typical company or mean-profit-revenue relation. In the `typical economy' picture 
discussed here, expected mean profits behave directly proportional to revenue, the latter being the 
driving variable. Furthermore, conjectures allow us to study profit fluctuations depending on actual 
revenue plus a fluctuating term governed by a distribution function of a L\'evy type. Strong deviations 
from the expected typical behavior of companies can be referred to as extremal, non-typical events. 
Indeed, within the 54 years data considered few extreme events (years) are observed for which profits 
and revenues display strong deviations from the linear relation between $P_0$ and $R_0$. However, such 
deviations can be traced back to specific historical facts and the corresponding years may be considered
as non-typical ones.

Although our conclusions are based on the study of the highest revenue companies in the U.S., we believe 
that our results still possess a robust degree of universality to be of more general validity, and 
can be used as a benchmark from which one can predict forthcoming profit-revenue scenarios. As an 
application, the present analysis has been used to estimate the probability that a single company 
has in order to become profitable. The analyzed empirical data are in support of our suggestions.

\acknowledgments
HER would like to thank Claudio Brighenti for illuminating discussions.



\end{document}